# Machine prediction of topological transitions in photonic crystals

Bei Wu[1,*], Kun Ding[2,3,*], C. T. Chan[2], and Yuntian Chen[1,4,★]

[1]*School of Optical and Electronic Information, Huazhong University of Science and Technology, Wuhan 430074, China*

[2]*Department of Physics and William Mong Institute of Nano Science and Technology, Hong Kong University of Science and Technology, Clear Water Bay, Kowloon, Hong Kong, China*

[3]*The Blackett Laboratory, Department of Physics, Imperial College London, London SW7 2AZ, United Kingdom*

[4]*Wuhan National Laboratory for Optoelectronics, Huazhong University of Science and Technology, Wuhan, China*

[*] These two authors contributed equally to this work.

[★] Email: yuntian@hust.edu.cn

**Abstract.** We train artificial neural networks to distinguish the geometric phases of a set of bands in 1D photonic crystals. We find that the trained network yields remarkably accurate predictions of the topological phases for 1D photonic crystals, even for the geometric and material parameters that are outside of the range of the trained dataset. Another remarkable capability of the trained network is to predict the boundary of topological transition in the parameter space, where a large portion of trained data in the vicinity of that boundary is excluded purposely. The results indicate that the network indeed learns the very essence of the structure of Maxwell's equations, and has the ability of predicting the topological invariants beyond the range of the training dataset.

*Introduction.* Recent achievements in machine learning including the neural network (NN) have triggered considerable attention of applying NN to solve practical engineering problems [1–6], as well as hard problems in the field of optics and photonics [7–15]. For example, inverse design problems in optics and photonics often require a lot of tedious optimization calculations and NN can be more efficient. Some recent work succeeded in using NN to achieve the reverse design of nanostructures, including the design of meta-surfaces [7], multilayer nanoparticles [8] and multilayer thin films [9]. In addition, NN has also been utilized to solve the inverse design problem of topological photonics [11,12]. These NN based works show higher accuracy and time-efficiency than that of conventional procedures. The high prediction accuracy can be interpreted as an advanced interpolation function implemented by NN, which works well within the parameter space of the training dataset. Such high accurate predictions can be of great help for the photonic design and engineering, if the high prediction accuracy can be extended beyond the training dataset.

A few attempts of predicting physical quantities beyond the training dataset have been made in the context of condense matter physics, with emphasis on the classifications of the topological order or phase transitions [16–25]. For instance, Zhang *et al.* managed to train NN to predict the larger winding numbers beyond the training data set in one-dimensional (1D) modified Su-Schrieffer-Heeger model, by feeding the NN with parameterized Hamiltonian [16]. The same concept is also extended to a four-band 1D problem, as well as a two-band 2D problem [17], where the associated topological invariants (Chern numbers) are examined. In addition, the ground state wave functions are also used directly as the training data to predict quantum phase transition [18,19]. Notably, the aforementioned works are mainly confined to simplified models, where the tight binding approximation is valid. It is interesting and relevant to ask whether the capability of extrapolating the topological order in condense matter physics system can be extended to photonic system that is beyond the tight binding approximation.

In this paper, we examine the predication capability of NN for photonic structures, with the input encoded with the structure of Maxwell's equations in momentum space. As a nontrivial example, we manage to use convolutional neural network (CNN) to predict the topological invariant of 1D photonic crystal (PC) for geometric configurations which lie outside the parameter space of the training dataset. Importantly, the CNN can detect the boundaries of the topological transitions in the parameter space.

*Photonic crystal and neural network.* As shown in Fig. 1(a), we consider an AB layered PC with the unit cell marked by yellow dotted lines. The dielectric constant of A(B) layer is $\varepsilon_a(\varepsilon_b)$, and the thickness of A(B) layer is $d_a(d_b)$. The lattice constant $\Lambda = d_a + d_b$ is fixed throughout this work. Figure 1(b) shows the dispersion of the lowest four bands (solid lines) calculated using the transfer matrix method (TMM) [26,27]. As the system has inversion symmetry, the geometric Zak phase [28] of each Bloch band is either 0 or π, with a corresponding winding number of 0 or 1. We calculate the Zak phases of the bands and label each band with its winding number. As for the lowest four bands considered in this paper, the four-binary-number labeling is translated into a decimal integer, *e.g.*, *1011* in Fig. 1(b) to *11* for convenience of the one-hot labeling of the output of the CNN.

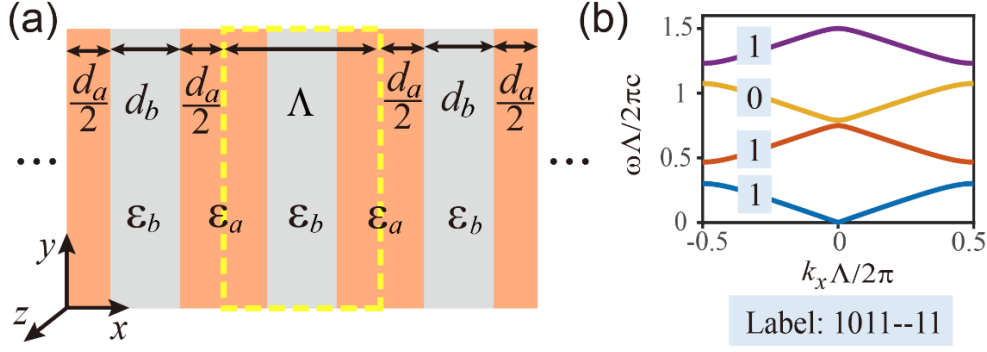

**Fig. 1** (a) Sketch of an AB layered photonic crystal; (b) Band dispersion and Zak phase labeling of the lowest four bands for the structure with $\varepsilon_a = 4$, $\varepsilon_b = 1$, $d_a = 0.3\Lambda$, $d_a = 0.7\Lambda$.

To train the NN to learn the underlying principle as governed by the Maxwell's equations as well as the structural and material characteristics of the PCs, we feed the CNN with the Hamiltonian of PC and mine the topological information from the input data so that it can potentially predict the topological invariants of PCs. The plane wave method (PWM) [29,30] is utilized to generate Hamiltonian of photonic crystals with $m_H$ plane waves used in the truncated basis. The value of $m_H = 61$ is carefully selected to balance the numerical errors of PWM as well as the size of input data of NN. Detailed comparison with TMM can be found in section II of Supplementary Materials [31]. With the 11 sampled Bloch $k$-vectors, *i.e.*, $n_K = 11$, the size of the input Hamiltonian for one PC within the Brillouin zone is $[n_K, m_H, m_H, col]$, where the last dimension $col = 2$ represents the real and imaginary parts of Hamiltonian.

Given the input Hamiltonians (consisting a rank-4 tensor $[n_K, m_H, m_H, col]$) and binary labelling, we proceed to study the architecture of CNN to realize the machine predication of topological transitions in PCs. As a simple example shown in Fig. 2, we consider the CNN workflow containing two convolutional layers (CLs) and two fully-connected layer (FLs). The two CLs contain $N_{C1}$ and $N_{C2}$ kernels of size [2,3,3] respectively, as well as a max-pooling operation with size [1,2,2], and the two FLs have $N_{F1}$ and $N_{F2}$ neurons with $N_{F2} = 16$ as the output layer has 16 neurons. The one-hot encoding output is a rank-1 tensor with shape [16], which has one-to-one correspondence to the binary labelling. For instance, *11* is represented as that only the 12th neuron of the output is *1* while others are *0*. In principle, more layers could be used to train the network to obtain a better performance.

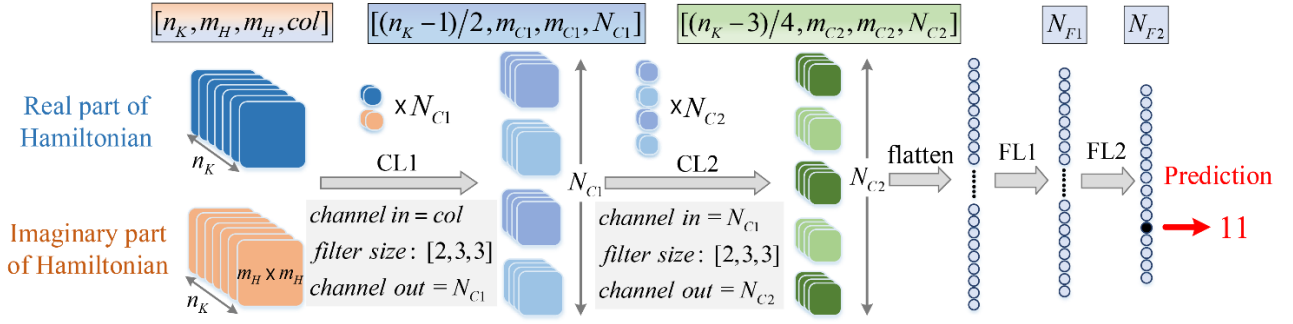

**Fig. 2** Schematic diagram of the CNN with two convolutional layers and two fully connected layers. The input end of CNN (left hand side) consists of the real and imaginary parts of Hamiltonians generated by PWM. The output end of CNN (right hand side) is the prediction of Zak phase sequence with one-hot encoding of four bands.

*Preparation of datasets and network training.* Next, we discuss the preparation of the datasets, the parameter space of which has three degrees of freedom (DOFs) $(\varepsilon_a, \varepsilon_b, d_a)$, as shown in Fig. 3(a). To guarantee the numerical accuracy of PWM, *i.e.*, avoiding the band crossing and small bandgap, two constraints to the range of the parameter space are imposed, *i.e.*, $\varepsilon_a \geq \varepsilon_b + 0.5$ and $0.1 \leq d_a/\Lambda \leq 0.9$. In Fig. 3(a), the parameter space is divided into four sectors, corresponding to four datasets labelled as *Train-1(Train-2)*, *Test-1*, *Test-2* and *Test-3*. The parameter range of each dataset is indicated in Fig. 3(a) (see Supplementary Materials, section I for details of the four datasets [31]). As an example, Fig.3(b) shows the distribution of the labelling in one training dataset, plotted as functions of $d_a$ and $\varepsilon_a$ for the $\varepsilon_b = 2$ plane within the *Train-1* dataset. The six colors of the solid circles represent the six labels defined in Fig. 1(b), corresponding to the following six cases *1–0001, 5–0101, 7–0111, 9–1001, 11–1011, 15–1111*. Each case has 600 randomly distributed samples and there are three planes ($\varepsilon_b = 1, 2, 3$) within *Train-1*, leading to 10800 (=600 × 6 × 3) samples in total for *Train-1* dataset. The five solid lines ($T_1 \sim T_5$) in Fig. 3(b) are the boundaries separating different labels determined using TMM. The five solid lines calculated using TMM perfectly match the borders of the six different colors calculated using PWM, showing that the TMM and PWM give consistent results of winding numbers for the lowest four bands. To examine the capability and possibility of our proposed CNN, we conceive a different training dataset, *i.e.*, *Train-2* dataset, which is identical to *Train-1* dataset except that a portion of samples in the neighborhood of the $T_3$ line (black dashed line) is omitted purposely as shown in Fig. 3(c). The fraction of the omitted data in Fig. 3(c) is $r = 0.5$, which is defined as the width of $rd_T$ of the omitted data divided by the difference of $d_a$ at $T_2$ and $T_4$ ($d_T$) for any $\varepsilon_a$.

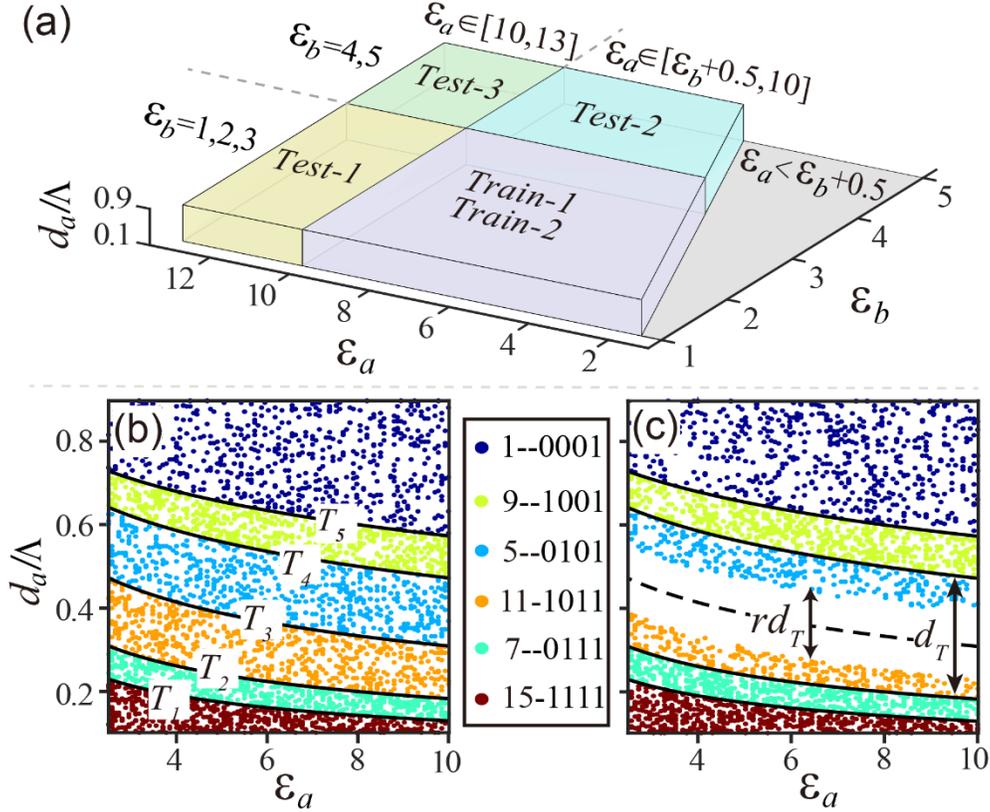

**Fig. 3** (a) Schematic representation of two training datasets and three test datasets, occupying different sectors of the parameter space; (b)/(c) The distribution of different topological phase sequences of samples in the $\varepsilon_b = 2$ plane within dataset *Train-1 /Train-2* with $r = 0.5$.

Once the CNN is set up, one just need to feed NN with the training data including both Hamiltonian and binary labelling. For self-testing purpose, each training dataset is separated into two subsets, 70% for training and 30% for testing. We follow the standard routine of supervised learning to train the network [32–34], *i.e.*, running the learning and testing procedure simultaneously and iteratively until the desired accuracy is fulfilled, which can be saved and used to predict topological transitions with parameters outside the training dataset.

***Predicting the topological transitions beyond the database.*** As the first example, we use the dataset *Train-1* to train the network, which consists of three CLs with $[N_{C1}, N_{C2}, N_{C3}] = [10,20,20]$, and three FLs with $[N_{F1}, N_{F2}, N_{F3}] = [300,100,16]$. The trained network is used to predict the boundaries of the parameter space (for examples, the $T_1 \sim T_5$ lines in Fig. 3(b)), where the topological transition occurs. For a given $\varepsilon_a$, the predicted phase transition points (the critical values $d_{cnn}$ at which the transition occurs) is obtained by scanning $d_a$ and taking the mean value of two adjacent $d_a$ with different Zak phases. To avoid numerical instability, such calculations of the trained CNN are repeated 10 times with different initialization conditions. We take the mean value $\overline{d_{cnn}}$ of these 10 $d_{cnn}$ as the predicted results, and the standard deviation is treated as error. The topological transition lines within *Train-1* dataset predicted by the well-trained CNN in the $\varepsilon_b = 2$ plane are shown by open circles with error bars in Fig. 4(a) with purple background. We see that the size of open circles can already cover the error bar, indicating that the predicted results obtained from the well-trained CNN are stable. For comparison, we also plot the topological transition lines calculated by TMM in the $\varepsilon_b = 2$ plane, as shown by solid lines in Fig. 4(a). The excellent agreement between TMM and CNN shows the well-trained CNN can reproduce the five topological phase transitions within the training dataset.

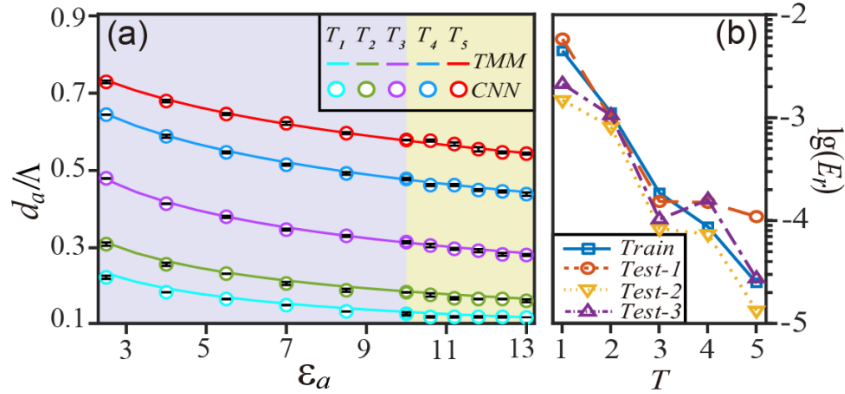

**Fig. 4** (a) The transition lines ($\varepsilon_b = 2$) within *Train-1* (purple background) and *Test-1* (yellow background) predicted by the well-trained CNN (circles) and TMM (open solid lines); (b) The relative errors $E_r$ between CNN prediction and TMM for the training dataset and three test datasets.

To demonstrate the predication capability of the network, we use the trained CNN by dataset *Train-1* to predict four-band-Zak-phase labeling and associated phase transitions in datasets *Test-1*, *Test-2* and *Test-3*. In doing so, we are extrapolating to parameter values outside the training datasets. Figure 4(a) with yellow background shows a comparison of the predicted transition line (open circles) and TMM results (solid lines) for constant $\varepsilon_b = 2$ within *Test-1*. The relatively low error of CNN results, which is less than $10^{-3}$, shows high predication accuracy and stability beyond the range of the training data. Evidently, the predication of dataset *Test-1*, *Test-2* and *Test-3* is not a simple interpolation of the topological invariant of *Train-1*, due to the fact that the *Test-1* dataset is outside the range of training data, as shown in Fig. 3(a). The remarkable agreement between TMM and CNN shows that our network indeed encodes some characteristics of Maxwell's equation from the input data and captures the essence of topological properties of photonic bands of different PCs.

To further evaluate the prediction accuracy of our network, we consider the relative prediction error of the topological transitions of our network with reference to TMM analytical results defined as $E_r = \frac{1}{N}\sum_{i=1}^{N}(\frac{|\overline{d_{cnn}}-d_{tmm}|}{d_{tmm}})^2$, where $N$ is the total number of data points in the phase transition surfaces (sets of $T_1 \sim T_5$ transition lines for different values of $\varepsilon_b$ corresponding to the light blue surfaces in Fig. S1 (see Supplementary Materials, section I [31]). The $T_i$ associated phase transition surface refers to the collection of the transition lines of $T_i$ in 3D parameter space $(d_a, \varepsilon_a, \varepsilon_b)$, with details given in section III of Supplementary Materials [31] and $d_{cnn}$ ($d_{tmm}$) corresponds to the topological transition point calculated by CNN (TMM). In Fig. 4(b), the relative errors for datasets *Train-1*, *Test-1*, *Test-2* and *Test-3* as a function of five transition surfaces are plotted by squares, circles, inverted triangles and triangles, respectively. Evidently, the predication of topological phase transitions is very accurate, with the mostly error less than $10^{-3}$. The prediction error $E_r$ of *Test-2* and *Test-3* test datasets are apparently less than that of *Test-1*. This somewhat surprising result is due to the well-known truncation errors intrinsic to the PWM photonic crystal with a large index contrast between its constituent components (see Supplementary Materials, section II [31]).

***Prediction of unknown transitions.*** We now examine the performance of the CNN trained using a different dataset *Train-2*, where the samples in the neighborhood of one of the transition lines ($T_3$) are removed. This width of the region of missing data is controlled by a parameter *r*, as illustrated by the white region in Fig. 3(c). The architecture of CNN used here is similar to but slightly more complex than the one trained by *Train-1*, and it contains three CLs with $[N_{c1}, N_{c2}, N_{c3}] = [24,32,48]$ and three FLs with $[N_{l1}, N_{l2}, N_{l3}] = [500,300,16]$. We train the CNN based on the *Train-2* dataset with $r = 0.2$ and apply it on the four datasets. Figure 5(a) shows the phase transition lines $T_1 \sim T_5$ predicted by CNN (open circles) and calculated by TMM (solid lines) in the $\varepsilon_b = 2$ plane within *Train-2*, and Fig. 5(b) depicts the results in the $\varepsilon_b = 4$ plane within *Test-2*. In Figs. 5(a) and 5(b), the purple (blue) background indicates the different range within the *Train-2* (*Test-2*) dataset and the white color indicate the parameter space with omitted data. Remarkably, the predicted transition lines associated with $T_1 \sim T_5$ using our network agree quite well with the theoretical ones by TMM, especially accurate predication of $T_3$ transition line highlighting the extrapolation capability of our network. As shown in Fig. 5(b), the predicted results for the other three test datasets are also fairly good. Figure 5(c) shows that the relative errors $E_r$ of the phase transition surfaces calculated by the trained CNN based on *Train-2* with $r = 0.2$ (see Supplementary Materials, section IV [31]). As expected, $E_r$ at $T_3$ is the largest, and the error is still quite small.

The overall extrapolation capability of our network can be seen in Fig. 5(d) from the relative error of the $T_3$ transition surface for the four datasets, as the fraction $r$ of the omitted training data varies. The relative errors associated with the predicted transition surfaces in *Test-2* and *Test-3* datasets are smaller than those in *Test-1*. As expected, the training error is relatively small for smaller value of $r$. As $r$ increases, the relative error of the overall predicted transition surface will increase. Thus, the extrapolation capability of the same CNN structure is compromised, due to the fact that more data are omitted.

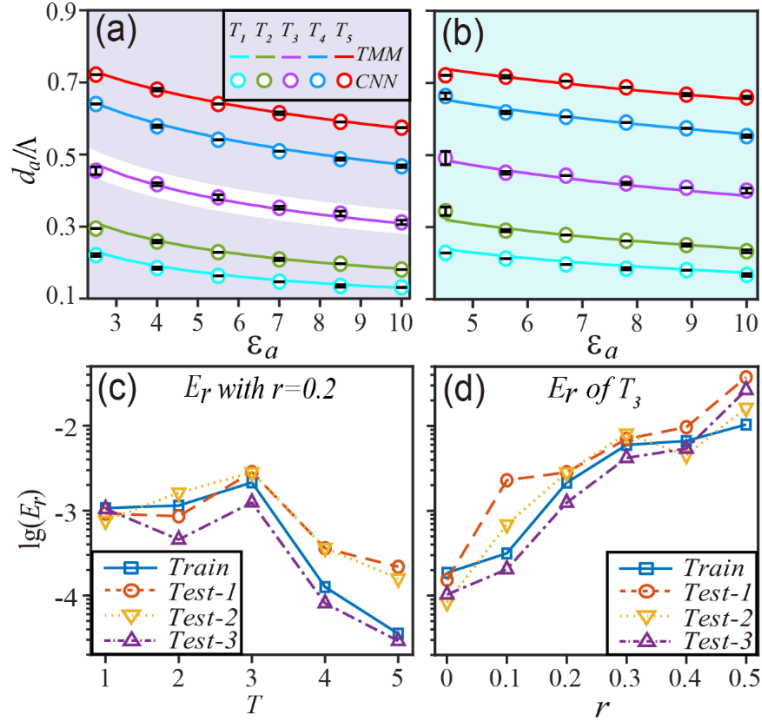

**Fig. 5** The topological phase transition lines for (a) $\varepsilon_b = 2$ in *Train-2* dataset with $r = 0.2$, (b) $\varepsilon_b = 4$ in *Test-2* dataset predicted by the well-trained CNN (trained by *Train-2*) are shown by open circles with error bars; (c) The relative error $E_r$ for the training dataset and the three test datasets; (d) The relative error $E_r$ for $T_3$ as a function of the fraction $r$ for all the four datasets.

***Discussion.*** Through the training of CNN and the prediction results of Fig. 4 and Fig. 5, we can see that CNN has an excellent interpolation ability and a reasonably good extrapolation capability in learning the critical conditions of geometric phase transition. Since we use the phase transitions predicted based on the Hamiltonian obtained by PWM to match that of the TMM, the error between PWM and TMM should be taken into account. In Fig. S3 (see Supplementary Materials, section II [31]), we can see that the relative error $E_r^p$ between PWM and TMM of the same dataset decreases from $T_1$ to $T_5$, which is in line with the trend of the error $E_r$ in Fig. 4(b). Moreover, the error $E_r^p$ is of the same order of magnitude as $E_r$ in Fig.

4(b), indicating that the error from CNN is small. We also note that the performance of *Test-2* and *Test-3* is better than that of *Test-1*. This is due to the fact that a larger contrast between $\varepsilon_a$ and $\varepsilon_b$ will bring a larger truncation error to the PWM for a fixed number of plane waves. It can be further seen that in the case where there is a boundary in the training set that does not specify a certain phase change, the performance of the CNN is still acceptable when $r$ is small. As the ratio $r$ of the omitted training set is larger, the information content will be fewer. For a higher $r$, the error of prediction will be larger for the same CNN architecture (see Supplementary Materials, section IV [31]). This problem can potentially be mitigated by using a more complex network structure.

Based on the above CNN analysis, we can see the good learning ability of CNN for topological phase transition in 1D photonic crystals. The network is basically performing a pattern recognition in $k$ space. As the Hamiltonian contains the structural and material information of PCs as well as the Fourier transformed Maxwell equation, its ability is not limited to the first four bands. It can be extended to topological phase transitions for higher bands and the technique can be applied to 2D PCs or other complex structures to predict more physical characters with higher efficiency and accuracy.

This work is supported by Natural National Science Foundation (NSFC) (Grant No. 11874026), and the Research Grants Council of Hong Kong (Grant No. AoE/P-02/12). K.D. acknowledges funding from the Gordon and Betty Moore Foundation.

Reference

[1] C. Szegedy, W. Liu, Y.-Q. Jia, P. Sermanet, S. Reed, D. Anguelov, D. Erhan, V. Vanhoucke, and A. Rabinovich, in *2015 IEEE Conference on Computer Vision and Pattern Recognition (CVPR)* (IEEE, Boston, MA, USA, 2015), pp. 1–9.

[2] K. He, X. Zhang, S. Ren, and J. Sun, in *2016 IEEE Conference on Computer Vision and Pattern Recognition (CVPR)* (IEEE, Las Vegas, NV, USA, 2016), pp. 770–778.

[3] D. Silver, A. Huang, C. J. Maddison, A. Guez, L. Sifre, G. van den Driessche, J. Schrittwieser, I. Antonoglou, V. Panneershelvam, M. Lanctot, S. Dieleman, D. Grewe, J. Nham, N. Kalchbrenner, I. Sutskever, T. Lillicrap, M. Leach, K. Kavukcuoglu, T. Graepel, and D. Hassabis, Nature **529**, 484 (2016).

[4] D. Silver, J. Schrittwieser, K. Simonyan, I. Antonoglou, A. Huang, A. Guez, T. Hubert, L. Baker, M. Lai, A. Bolton, Y. Chen, T. Lillicrap, F. Hui, L. Sifre, G. van den Driessche, T. Graepel, and D. Hassabis, Nature **550**, 354 (2017).

[5]  A. Graves, A. Mohamed, and G. Hinton, in *2013 IEEE International Conference on Acoustics, Speech and Signal Processing* (2013), pp. 6645–6649.

[6]  C. Chen, A. Seff, A. Kornhauser, and J. Xiao, in *2015 IEEE International Conference on Computer Vision (ICCV)* (IEEE, Santiago, Chile, 2015), pp. 2722–2730.

[7]  Z. Liu, D. Zhu, S. P. Rodrigues, K.-T. Lee, and W. Cai, Nano Lett. **18**, 6570 (2018).

[8]  J. Peurifoy, Y. Shen, L. Jing, Y. Yang, F. Cano-Renteria, B. G. DeLacy, J. D. Joannopoulos, M. Tegmark, and M. Soljačić, Sci. Adv. **4**, (2018).

[9]  D. Liu, Y. Tan, E. Khoram, and Z. Yu, ACS Photonics **5**, 1365 (2018).

[10]  I. Malkiel, M. Mrejen, A. Nagler, U. Arieli, L. Wolf, and H. Suchowski, Light: Science & Applications **7**, (2018).

[11]  L. Pilozzi, F. A. Farrelly, G. Marcucci, and C. Conti, Communications Physics **1**, (2018).

[12]  Y. Long, J. Ren, Y. Li, and H. Chen, Appl. Phys. Lett. **114**, 181105 (2019).

[13]  Y. Shen, N. C. Harris, S. Skirlo, M. Prabhu, T. Baehr-Jones, M. Hochberg, X. Sun, S. Zhao, H. Larochelle, D. Englund, and M. Soljačić, Nature Photon **11**, 441 (2017).

[14]  Y. Qu, L. Jing, Y. Shen, M. Qiu, and M. Soljačić, ACS Photonics **6**, 1168 (2019).

[15]  T. Yan, J. Wu, T. Zhou, H. Xie, F. Xu, J. Fan, L. Fang, X. Lin, and Q. Dai, Phys. Rev. Lett. **123**, (2019).

[16]  P. Zhang, H. Shen, and H. Zhai, Phys. Rev. Lett. **120**, 066401 (2018).

[17]  N. Sun, J. Yi, P. Zhang, H. Shen, and H. Zhai, Phys. Rev. B **98**, 085402 (2018).

[18]  T. Ohtsuki and T. Ohtsuki, J. Phys. Soc. Jpn. **85**, 123706 (2016).

[19]  T. Ohtsuki and T. Ohtsuki, J. Phys. Soc. Jpn. **86**, 044708 (2017).

[20]  E. P. L. van Nieuwenburg, Y.-H. Liu, and S. D. Huber, Nat. Phys. **13**, 435 (2017).

[21]  Y. Zhang and E.-A. Kim, Phys. Rev. Lett. **118**, 216401 (2017).

[22]  G. Carleo and M. Troyer, Science **355**, 602 (2017).

[23]  C. Wang, H. Zhai, and Y.-Z. You, ArXiv:1901.11103 [Cond-Mat, Physics:Quant-Ph] (2019).

[24]  D.-L. Deng, X. Li, and S. Das Sarma, Phys. Rev. B **96**, (2017).

[25] D.-L. Deng, X. Li, and S. Das Sarma, Phys. Rev. X **7**, (2017).

[26] A. Yariv and P. Yeh, *Optical Waves in Crystals: Propagation and Control of Laser Radiation* (Wiley, New York, 1984).

[27] M. Xiao, Z. Q. Zhang, and C. T. Chan, Phys. Rev. X **4**, 021017 (2014).

[28] J. Zak, Phys. Rev. Lett. **62**, 4 (1989).

[29] K. M. Ho, C. T. Chan, and C. M. Soukoulis, Phys. Rev. Lett. **65**, 3152 (1990).

[30] K. Sakoda, *Optical Properties of Photonic Crystals*, 2nd ed (Springer, Berlin ; New York, 2005).

[31] see Supplementary Materials at *link* for more computational details.

[32] N. Buduma and N. Locascio, *Fundamentals of Deep Learning: Designing next-Generation Machine Intelligence Algorithms*, First edition (O'Reilly Media, Sebastopol, CA, 2017).

[33] Y. LeCun, Y. Bengio, and G. Hinton, Nature **521**, 436 (2015).

[34] A. Géron, *Hands-on Machine Learning with Scikit-Learn and TensorFlow: Concepts, Tools, and Techniques to Build Intelligent Systems*, First edition (O'Reilly, Beijing Boston Farnham Sebastopol Tokyo, 2017).